\documentclass[epj]{svjour}
\usepackage{graphicx}
\usepackage{amssymb}

\begin{document}

\title{Phonon-assisted phase separation in strongly correlated systems}
\author{Alexei Sherman}%
\institute{Institute of Physics, University of Tartu, W. Ostwaldi Str 1, 50411 Tartu, Estonia\\
\email{alekseis@ut.ee}}

\date{Received: date / Revised version: date}

\abstract{We relate the phase separation observed in many crystals with pronounced electron correlations to the regions of negative electron compressibility. They were found in several models describing strong electron correlations. At low temperatures, these regions arise near chemical potentials corresponding to the change of the ground state in the site Hamiltonian. The negative electron compressibility leads to the separation of the system into electron-rich and electron-poor domains. The energy released in the course of this separation is absorbed by phonons. Another role of phonons is to give a definite form -- stripes or checkerboards -- to lattice distortions and domains of different electron concentrations. The shape, direction, and periodicity of such textures are determined by wave vectors of lattice distortions, which most strongly scatter electrons.
\PACS{{xx.xx.xx}{xx}}
}

\maketitle

\section{Introduction}
The phase separation (PS) observed in many crystals with pronounced electron correlations is characterized by the creation of two or more phases with different electron concentrations. Other manifestations of the PS are the inhomogeneity of equilibrium positions of lattice distortions and spin density waves. Considerable attention was paid to the stripe order first observed in yttrium and lanthanum cuprate perovskites using neutron scattering \cite{Tranquada88,Birgeneau}. The structure of stripes was revealed in the nickelate La$_2$NiO$_{4+\delta}$ \cite{Tranquada94} and Nd-substituted cuprate La$_{2-x-y}$Nd$_y$ Sr$_x$Cu$_2$O$_{4+\delta}$ \cite{Tranquada95} as one-dimensional regions with excess holes separated by antiferromagnetic domains. Later, the PS was found in other cuprate perovskites \cite{Kremer,Hanaguri,Vershinin}, manganites \cite{Renner,Rao}, iron pnictides \cite{Park,Lang}, and mixed-valent rare-earth cobaltates \cite{Shenoy}. In some of these crystal families, shapes of phase-separated domains differ from stripes \cite{Rao,Park}. The participation of lattice distortions in the PS formation manifests itself in anomalies of phonon spectra \cite{Pintschovius,Reznik,Tacon}. The pronounced softening of phonon frequencies at wave vectors characterizing the PS spacial periodicity indicates lattice distortions, which are strong enough for the occurrence of anharmonic effects. All these experimental data points to the fact that the PS is the common property of crystals with pronounced electron correlations.

The first theoretical works \cite{Hizhnyakov,Machida,Zaanen} devoted to the PS in cuprates used the mean-field approximation for the solution of the $t$-$J$ or Hubbard models describing strongly interacting electrons. The PS mechanism suggested in these works was connected with the kinetic energy gain, which electron acquires in the ferromagnetic environment in comparison with the antiferromagnetic background. If this gain exceeds the energy cost of the distortion of the spin antiferromagnetic ordering, there appears an elementary excitation termed ferron -- an electron or hole surrounded by a ferromagnetic cluster. Previously, similar excitations were considered in soft magnetic semiconductors \cite{Nagaev}. In cuprates, it was supposed that ferromagnetic clusters overlap at a certain concentration of carriers, which leads to percolative conductivity in the crystal. This picture resembles experimental observations. However, more exact calculations using the two-dimensional (2D) $t$-$J$ model and the spin-wave approximation \cite{Sabczynski} did not confirm the formation of large enough ferrons for parameters of cuprates. The subsequent works applied more elaborate methods -- Monte-Carlo simulations, the Gutzwiller approximation, density matrix renormalization group, and strong coupling diagram technique (see, e.g., \cite{Moreo,Seibold,Sherman08,Chia,Heiselberg,White}). In some of these works, the PS was obtained in systems of interacting electrons. Authors of other investigations concluded that the electron-phonon interaction is an essential part of the PS mechanism. As a consequence, at present, there is no commonly excepted view on the PS mechanism in the mentioned crystals.

The common property of all these crystals is the strong electron correlations, and we relate the origin of the PS to this property. The correlations lead to the occurrence of the regions of negative electron compressibility (NEC) in the dependence of the electron concentration on the chemical potential, $x(\mu)$. At low temperatures, such regions appear near values of $\mu$ corresponding to the change of the ground state of the site Hamiltonian \cite{Sherman20a,Sherman20b}. The NEC regions arise due to sharp variations of bandwidths and related significant variations of electron correlations, which are concomitant to the ground-state change. For example, the NEC arises when a sharp reduction of the bandwidth occurring with increasing $x$ leads to a decrease of $\mu$. As a result, the electron compressibility $\kappa=x^{-2}({\rm d}x/{\rm d}\mu)$ becomes negative. For the case of positive $\kappa$, deviations of $x$ from a mean value are energetically unfavorable. For $\kappa<0$, it is energetically favorable to separate the system into electron-rich and electron-poor domains. It takes place if there exists a subsystem, which is able to absorb the released energy. In solids, such a subsystem is lattice vibrations. Peculiarities of the phonon spectrum and the electron-phonon interaction constants determine the spatial distribution of the electron domains and respective lattice distortions.

\section{Negative electron compressibility}
In the case of strong electron correlations, it is reasonable to apply the perturbation series expansion around the atomic limit. Apparently, for the first time, this idea was suggested by Hubbard in \cite{Hubbard}. Later on, it was realized in works using the diagram technique for Hubbard operators \cite{Zaitsev,Izyumov,Ovchinnikov} and the strong coupling diagram technique (SCDT) \cite{Vladimir,Metzner,Pairault,Sherman18,Sherman19a,Sherman19b}. The latter approach is based on a regular series expansion of Green's functions in powers of hopping constants in the electron kinetic energy of the Hamiltonian. Terms of the series are products of these constants and on-site cumulants of electron creation and annihilation operators. For the 2D one-band repulsive Hubbard model, the validity of the SCDT was proved in comparison with the results of numeric experiments and experiments with ultracold fermionic atoms in 2D optical lattices \cite{Sherman18,Sherman19a,Sherman19b}. In particular, it was shown that the critical repulsion for the Mott metal-insulator transition is close to that observed in Monte-Carlo simulations. For the comparable parameters, spectral functions and densities of states are similar to those found in exact diagonalizations and Monte Carlo simulations. Temperature and concentration dependencies of the uniform spin susceptibility, spin structure factor, square of the site spin, and double occupancy are in good agreement with results of Monte Carlo simulations, numeric linked-cluster expansion, and experiments with ultracold fermionic atoms. Shapes and intensity distributions in Fermi surfaces in electron- and hole-doped cases are similar to those observed experimentally. Lastly, moments sum rules are fulfilled with good accuracy.

\begin{figure}[t]
\centerline{\resizebox{0.99\columnwidth}{!}{\includegraphics{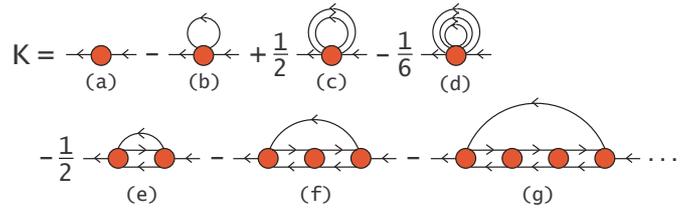}}}
\caption{Diagrams of several lowest orders in the SCDT expansion for $K({\bf k},j)$.} \label{Fig1}
\end{figure}

The linked-cluster theorem is valid, and partial summations are allowed in the SCDT applied to the Green's function $G({\bf l'}\tau';{\bf l}\tau)=\langle{\cal T}a^\dagger_{\bf l'\sigma}(\tau')a_{\bf l\sigma}(\tau)\rangle$. Here $a^\dagger_{\bf l\sigma}$ and $a_{\bf l\sigma}$ are the creation and annihilation electron operators on the lattice site {\bf l} with the spin projection $\sigma=\pm 1$, time dependencies and the thermodynamic averaging are determined by the system Hamiltonian, and ${\cal T}$ is the chronological operator (hereinafter, for the notation simplicity, we consider one electron band and one phonon branch; the generalization to several bands and branches is straightforward). As in the conventional diagram technique with the expansion in powers of an interaction \cite{Abrikosov}, the SCDT diagrams for the one-particle Green's function are divided into reducible and irreducible. In contrast to the former, the latter cannot be separated into two disconnected parts by cutting a hopping line. If the sum of all irreducible diagrams -- the irreducible part -- is denoted by $K({\bf k},j)$, the Fourier transform of the Green's function can be written as
\begin{eqnarray}\label{Larkin}
G({\bf k},j)&=&K({\bf k},j)+K({\bf k},j)t_{\bf k}K({\bf k},j)\nonumber \\
  &&\quad+K({\bf k},j)t_{\bf k}K({\bf k},j)t_{\bf k}K({\bf k},j)+\ldots\nonumber \\
  &=&\big\{[K({\bf k},j)]^{-1}-t_{\bf k}\}^{-1},
\end{eqnarray}
where {\bf k} is the wave vector, $j$ is the integer defining the Matsubara frequency $\omega_j=(2j-1)\pi T$ with the temperature $T$, and $t_{\bf k}$ is the Fourier transform of hopping constants. Several lowest order diagrams for $K$ are shown in figure~\ref{Fig1}.

In this figure, circles are cumulants \cite{Kubo} of electron operators. The cumulant order is determined by the number of incoming (outgoing) lines. The first-order cumulant is given by the expression $C^{(1)}(\tau',\tau)=\langle{\cal T}a^\dagger_\sigma(\tau')a_\sigma(\tau) \rangle_0$, where the subscript 0 near angle brackets indicates that time dependencies and the thermodynamic averaging is determined by a site Hamiltonian $H_{\bf l}$. Usually, it consists of terms of the system Hamiltonian describing interactions inside a crystal unit cell. In this case, due to the translation symmetry, the cumulant does not depend on the site index. Therefore, it is omitted in the above formula. However, in the SCDT, some cluster can be also considered as a local system, and the SCDT power expansion is carried over hopping terms between such clusters (see, e.g., \cite{Senechal}). For the one-band Hubbard model with the Hamiltonian
\begin{equation}\label{Hubbard}
H=\sum_{\bf l'l\sigma}t_{\bf ll'}a^\dagger_{\bf l'\sigma}a_{\bf l\sigma}+\sum_{\bf l}H_{\bf l},
\end{equation}
the site Hamiltonian reads
\begin{equation}\label{site}
H_{\bf l}=\sum_\sigma\bigg(\frac{U}{2}n_{\bf l\sigma}n_{\bf l,-\sigma}-\mu n_{\bf l\sigma}\bigg),
\end{equation}
where $t_{\bf ll'}$ is the hopping constant, $U$ is the on-site Co\-u\-lomb interaction, and the number operator $n_{\bf l\sigma}=a^\dagger_{\bf l\sigma}a_{\bf l\sigma}$. For calculating cumulants, it is convenient to use the representation of eigenvectors $|\lambda\rangle$ of the site Ha\-mil\-to\-ni\-an and the generalization of the Wick theorem for respective Hubbard operators \cite{Zaitsev,Izyumov,Ovchinnikov}. For the Fourier transform of the first-order cumulant the result reads
\begin{equation}\label{C1}
C^{(1)}(j)=\frac{1}{Z}\sum_{\lambda\lambda'}\frac{{\rm e}^{-\beta E_\lambda}+{\rm e}^{-\beta E_{\lambda'}}}{{\rm i}\omega_j+E_\lambda-E_{\lambda'}}
\langle\lambda|a_\sigma|\lambda'\rangle \langle\lambda'|a^\dagger_\sigma|\lambda\rangle,
\end{equation}
where $E_\lambda$ is an eigenenergy of $H_{\bf l}$ and the partition function $Z=\sum_\lambda\exp(-\beta E_\lambda)$ with $\beta=1/T$.

The expression for the second-order cumulant
\begin{eqnarray*}
&&C^{(2)}(\tau_1,\sigma_1;\tau_2,\sigma_2;\tau_3,\sigma_3;\tau_4,
\sigma_4)\\
&&\quad=\big\langle{\cal T}a^\dagger_{\sigma_1}(\tau_1)a_{\sigma_2}(\tau_2) a^\dagger_{\sigma_3}(\tau_3)a_{\sigma_4}(\tau_4)\big\rangle_0\\
&&\quad\quad -C^{(1)}(\tau_1,\tau_2)C^{(1)}(\tau_3,\tau_4)\delta_{\sigma_1 \sigma_2}\delta_{\sigma_3\sigma_4}\\
&&\quad\quad+C^{(1)}(\tau_1,\tau_4)C^{(1)}(\tau_3, \tau_2)\delta_{\sigma_1 \sigma_4}\delta_{\sigma_3 \sigma_2}
\end{eqnarray*}
is more complicated. It can be found in \cite{Sherman20b}. Equations for both cumulants, as well as for cumulants of higher orders, contain the Boltzmann factors $\exp(-\beta E_\lambda)$. Due to these factors, at low temperatures, the ground state (or, in the case of degeneracy, states) and states obtained from it (them) by the creation or annihilation of an electron make the main contribution to the cumulants and, through equation (\ref{Larkin}), to Green's function. The ground state(s) of the site Hamiltonian is (are) changed with $\mu$ due to the term $-\mu n$ in the eigenenergies $E_\lambda$. Here $n$ is the number of electrons in the eigenstate $|\lambda\rangle$. Hence, at low $T$, the cumulants and Green's functions are sharply changed at values of $\mu$ corresponding to level crossing -- points, in which one ground state (states) is (are) substituted with another one (other ones). For the Hamiltonian (\ref{site}), such values of the chemical potential are $\mu=0$ and $\mu=U$. For $\mu<0$, $0<\mu<U$, and $\mu>U$ the ground states of $H_{\bf l}$ are, sequentially, the empty $|0\rangle$, doubly degenerate singly occupied $|\sigma\rangle$, and the doubly occupied $|2\rangle$ states. Electron bands related to these states differ significantly in their properties, which is reflected in their widths and locations -- for $\mu<0$ and $\mu>U$ the bandwidths are close to uncorrelated ones, while for $0<\mu<U$ they are substantially reduced. As a consequence, at low temperatures, the dependence $x(\mu)$ appears as follows: far below $\mu=0$ the concentration grows monotonously with $\mu$; as the value 0 is approached, the band is squeezed and shifted to higher energies, which leads to a {\sl decrease} of $x$ with increasing $\mu$. It is the NEC region. At a further increase of $\mu$, the concentration starts to grow again, it levels out in the Mott gap and resumes growth till $\mu\approx U$ where the second NEC is located (due to the particle-hole symmetry, the dependence $x(\mu)$ has the center of symmetry in the point $x=1$, $\mu=U/2$).

\begin{figure}[t]
\centerline{\resizebox{0.99\columnwidth}{!}{\includegraphics{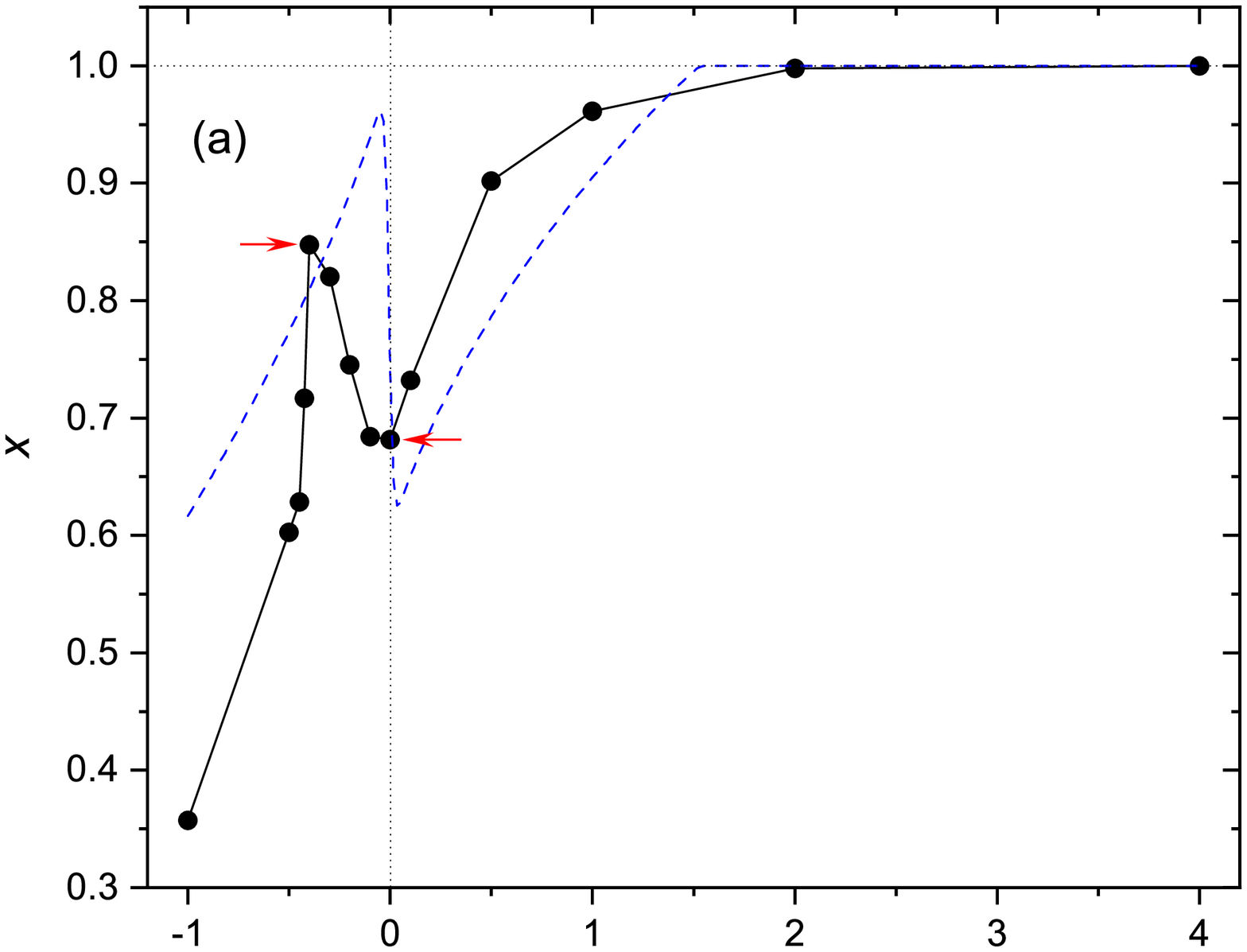}}}
\vspace{3ex}
\centerline{\resizebox{0.99\columnwidth}{!}{\includegraphics{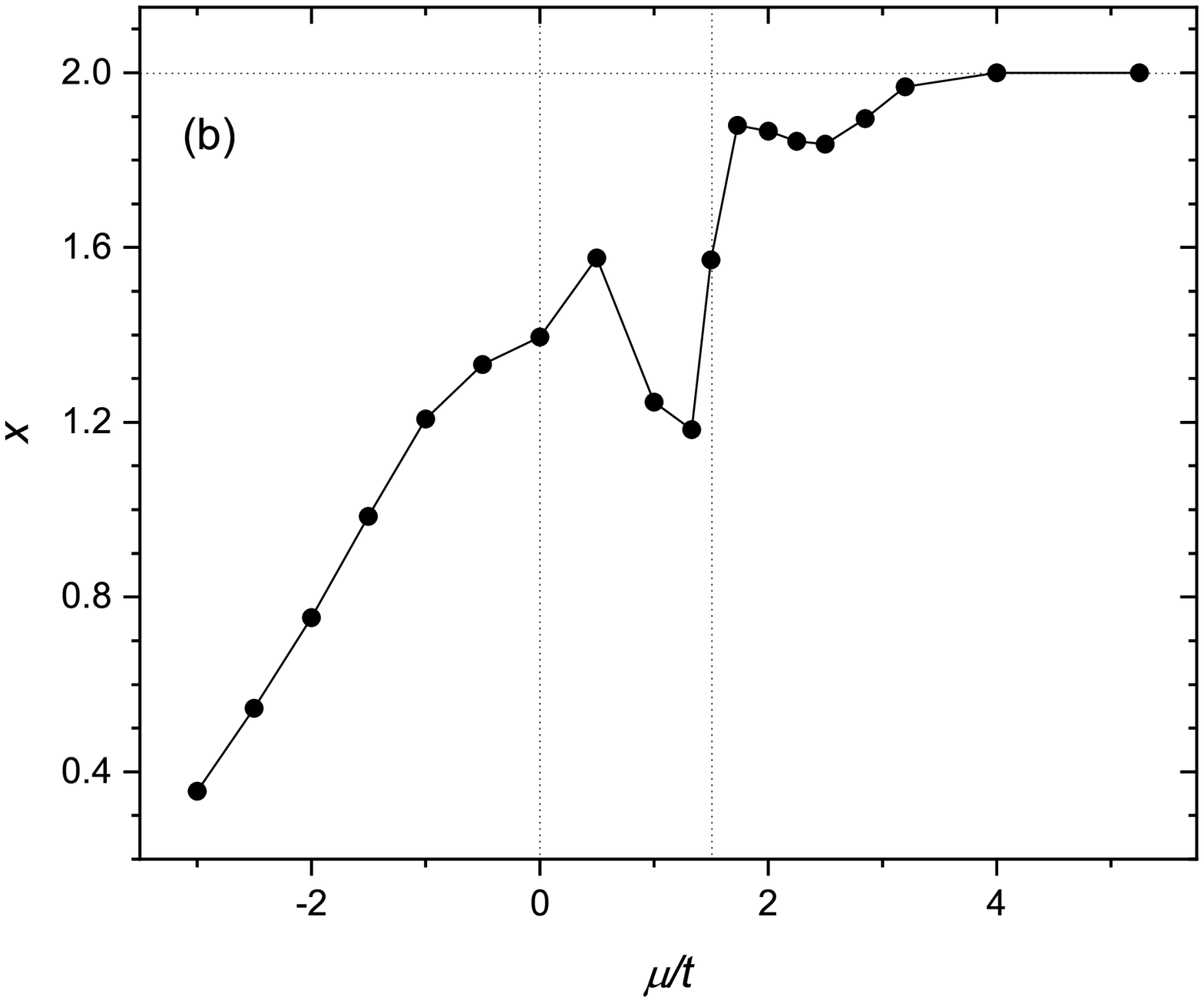}}}
\caption{Dependencies of the electron concentration $x$ on the chemical potential $\mu$ in the range from negative values to half-filling, $t$ is the hopping constant between nearest-neighbor sites. (a) The one-band 2D Hubbard model with the on-site Coulomb repulsion $U=8t$ and the temperature $T=0.13t$ (black circles and line) \protect\cite{Sherman20a}. The same dependence obtained in the Hubbard-I approximation for $T=0.01t$ (blue dashed line). (b) The two-orbital 2D Hubbard-Kanamori model with the repulsion between electrons on the same orbital $U=6t$, the exchange integral $J=1.5t$, and $T=0.13t$ \protect\cite{Sherman20b}.}  \label{Fig2}
\end{figure}
This dependence is shown in figure~\ref{Fig2}(a) by black circles. The result was obtained using the SCDT, in which infinite sequences of diagrams describing interactions of electrons with spin and charge fluctuations of all ranges were taken into account \cite{Sherman20a}. The width and depth of the NEC region depend on temperature. Let us denote the largest and smallest concentrations in the NEC as $x_r$ and $x_p$, respectively, and the corresponding values of the chemical potential as $\mu_r$ and $\mu_p$ (these points are indicated by red arrows in the figure). As $T$ increases from $T=0.13t$, the difference $\mu_r-\mu_p$ grows, while $x_r-x_p$ decreases. Finally, near $T=0.3t$ the NEC region disappear. On the contrary, as the temperature decreases to $0.01t$, $\mu_r-\mu_p$ decreases and $x_r-x_p$ grows, as shown in figure~\ref{Fig2}(a) by the blue dashed line. This curve is illustrative -- it was calculated in the Hubbard-I approximation \cite{Vladimir} since with all processes taken into account in \cite{Sherman20a} we had not yet been able to attain such low $T$. Below, this result will be used for estimating hole concentration in stripes.

To demonstrate the fact that NEC regions arise in other systems of strongly correlated electrons also, the dependence $x(\mu)$ in the 2D two-orbital Hubbard-Kanamori model is given in figure~\ref{Fig2}(b) \cite{Sherman20b}. The model includes the on-site Coulomb repulsions between electrons on the same and different orbitals as well as the spin-flip and pair-hopping terms characterized by Hund's coupling \cite{Georges}. In contrast to the one-band Hubbard model, this more complicated system has four critical chemical potentials corresponding to the level crossings in the site Hamiltonian. Two of them are indicated by dotted vertical lines in the figure. As in the case of the one-band Hubbard model, for low $T$, there exist NEC regions near these $\mu$, which are somewhat shifted to higher energies due to strong correlations. Results in figure~\ref{Fig2}(b) were also obtained using the SCDT with the summation of infinite sequences of diagrams describing interactions of electrons with spin, charge, and orbital fluctuations of all ranges in an infinite crystal \cite{Sherman20b}.

The results shown in figure~\ref{Fig2} demonstrate that NEC regions are inherent in the case of strong electron correlation when the change of the ground state of the site Hamiltonian leads to the substantial reconstruction of the whole electron spectrum.

\section{Phase separation}
Let us consider two crystal domains, in one of which, due to fluctuations, the electron concentration is slightly larger than the average one, $\bar{x}+\delta x$, $\delta x>0$, and in the other slightly smaller, $\bar{x}-\delta x$. The Helmholtz free energy $F$ satisfies the equation $\mu=\partial F/\partial (xN)|_{T,V}$, where $N$ and $V$ are the number of sites and volume, respectively. Therefore, the deviation of the free energy from the mean value is
\begin{equation}\label{Helmholtz}
\delta F/N=[\mu(\bar{x}+\delta x/2)-\mu(\bar{x}-\delta x/2)]\,\delta x.
\end{equation}
For the case of positive compressibility, this quantity is positive. Therefore, the system tends to eliminate the fluctuation of $x$. In the case of the NEC, $\delta F<0$ and it is energetically favorable to increase the difference of concentrations in the domains. The process of the charge separation will continue until $x$ in the electron-rich domain reaches the largest concentration in the NEC, $x_r$, and the electron-poor domain attains its smallest concentration, $x_p$. The domains are assumed to be large enough to use the dependence $\mu(x)$ calculated in a large crystal. Below, we suppose the perfect screening of all components of the Coulomb interaction except the on-site one and neglect the surface energy between two phases. This energy arises due to the size quantization of electron levels in them \cite{Balian,Sboychakov}. These two energy contributions influence on domain shapes. However, as follows from above experimental results, they are mainly determined by the electron-phonon interaction considered below. Without these contributions, electron-rich and electron-poor domains of any shapes and any connectedness with the number of sites $N_r=N(\bar{x}-x_p)/(x_r-x_p)$ and $N_p=N(x_r-\bar{x})/(x_r-x_p)$, respectively, provide the lowest electron free energy $F_e=F_r+F_p$.

In the above discussion, it is tacitly assumed that the energy released in the course of the PS into the two types of domains is absorbed by other parts of the system, such as phonons. To describe their influence, we add the following terms to the Hamiltonian (\ref{Hubbard}), (\ref{site}):
\begin{equation}\label{ev}
H_{ev}=\sum_{\bf kk'\sigma}\frac{v_{\bf kk'}}{\sqrt{N}}a^\dagger_{\bf k+k',\sigma}a_{\bf k\sigma} \big(b_{\bf k'}+ b^\dagger_{-\bf k'}\big)+\sum_{\bf k}\omega_{\bf k}b^\dagger_{\bf k}b_{\bf k},
\end{equation}
where the first term in the right-hand side describes the electron-phonon interaction with the constant $v_{\bf kk'}$ and phonon creation and annihilation operators $b^\dagger_{\bf k}$ and $b_{\bf k}$. The second term is the phonon energy with the phonon frequency $\omega_{\bf k}$. Terms of such type are frequently used in the description of the electron-phonon interaction in different crystals, and we consider them first in this section. However, it should be noted that for the cases of one-layer cuprates and out-of-plane oxygen vibrations other expressions for the interaction term and phonon energy have to be used. This case will be considered at the end of the section. The interaction constants $v_{\bf kk'}$ differ slightly in domains with distinct electron concentrations. To simplify the below consideration, we neglect this difference and use the same constant for both types of domains.

We aim to find equilibrium positions of lattice vibrations and locations of electron-rich domains minimizing the free energy of the combined electron-phonon system. These parameters are directly detected by the quasi-elastic neutron and X-ray scattering \cite{Tranquada88,Birgeneau,Comin}. In strongly correlated systems, the on-site Hubbard repulsion is much larger than the electron-phonon interaction. Otherwise, this interaction would be comparable to the electron bandwidth, which led to the formation of strong-coupling polarons with masses much larger than those observed. The\-re\-fo\-re, the influence of the electron-phonon interaction on the electron subsystem is moderate, and, as before, we can consider it as phase-separated into electron-rich and electron-poor domains. We shall find their locations and shapes, as well as respective lattice equilibrium positions, by calculating the effective action of the phonon subsystem in the electron-phonon system. It is described by the Hamiltonian ${\cal H}$ with terms (\ref{Hubbard}), (\ref{site}), and (\ref{ev}). The partition function of the system reads
\begin{eqnarray}\label{Z}
{\cal Z}&=&{\rm Tr}\big({\rm e}^{-\beta{\cal H}}\big)={\rm e}^{-\beta\Omega}\nonumber\\
&=&\int{\cal D}(\xi^*\xi q)\exp\bigg\{-\int_0^\beta{\rm d}\tau\bigg[\sum_{\bf k\sigma}\xi^*_{\bf k\sigma}\frac{{\rm d}}{{\rm d}\tau}\xi_{\bf k\sigma}+H(\xi^*,\xi)\nonumber\\
&&\;+\frac{1}{2}\sum_{\bf k}\bigg(m\bigg|\frac{{\rm d}q_{\bf k}}{{\rm d}\tau}\bigg|^2+w_{\bf k}\,q_{\bf k}q_{\bf -k}\bigg)\nonumber\\
&&\;+\sum_{\bf kk'\sigma}\sqrt{\frac{2m\omega_{\bf k'}}{N}}v_{\bf kk'}\,\xi^*_{\bf k+k',\sigma}\xi_{\bf k\sigma}q_{\bf k'}\bigg]\bigg\},
\end{eqnarray}
where $\Omega(T,V,\mu)$ is the Landau thermodynamic potential, $\xi^*_{\bf k\sigma}$ and $\xi_{\bf k\sigma}$ are Grassmann numbers defining electron coherent states, $q_{\bf k}$ is the Fourier transform of lattice distortions, $m$ is the atomic mass, and the dynamic constant $w_{\bf k}=m\omega_{\bf k}^2$. To simplify notations, we do not explicitly indicate time dependencies of the variables $\xi^*_{\bf k\sigma}$, $\xi_{\bf k\sigma}$, and $q_{\bf k}$. The continual integration in (\ref{Z}) is performed with the antiperiodic boundary conditions for Grassmann variables, $\xi_{\bf k\sigma}(\beta)=-\xi_{\bf k\sigma}(0)$, and the periodic conditions for complex variables, $q_{\bf k}(\beta)=q_{\bf k}(0)$ \cite{Negele}.

To obtain the effective vibration Lagrangian, let us integrate out electronic variables. The first two terms in the action in (\ref{Z}) give the electron partition function ${\cal Z}_e$ and Landau potential $\Omega_e$. They can be calculated from the relation $xN=\partial\Omega_e/\partial\mu|_{T,V}$,
\begin{eqnarray}\label{Omega}
\Omega_e/N&=&\int_{\mu_0}^\mu x(\mu')\,{\rm d}\mu'\nonumber\\
&=&\frac{2T}{N}\int_{\mu_0}^\mu\sum_{{\bf k}j}{\rm e}^{{\rm i}\omega_j\eta}G({\bf k},j)\,{\rm d}\mu',\quad\eta\rightarrow+0 \nonumber\\
&=&-\frac{2}{N\pi}\int_{\mu_0}^\mu\int_{-\infty}^{\infty}\sum_{\bf k}\frac{{\rm Im}\,G({\bf k},\omega)}{\exp(\beta\omega)+1}\,{\rm d}\mu'{\rm d}\omega,
\end{eqnarray}
where $G({\bf k},\omega)$ is the electron Green's function on the real frequency axis. The chemical potential $\mu_0$ is arbitrary and corresponds to the chosen zero of $\Omega_e$. In particular, $\mu_0$ can be selected so small that $x\approx 0$. The drawback of equation~(\ref{Omega}) in comparison with the usual diagrammatic expansion of the Landau potential \cite{Abrikosov} is the need for calculating Green's function for a large number of chemical potentials. The substantial advantage of this equation over the usual method is the possibility to use partial summations in an infinite series of diagrams representing Green's function. Because of comparatively simple and fast SCDT calculations, in our opinion, this advantage of equation (\ref{Omega}) prevails over its drawback. Examples of such calculations of Green's function for a large number of $\mu$ values are given in figure~\ref{Fig2}. As indicated above, the lowest value of the electron free energy is achieved in the PS state, $F_e=F_r+F_p$, where $F_r=\Omega_r+x_rN_r\mu_r$ and $F_p=\Omega_p+x_pN_p\mu_p$. In calculating $\Omega_r$ and $\Omega_p$ from (\ref{Omega}), Green's functions obtained for respective values of the chemical potential are applied.

To integrate out electronic variables in other terms of ${\cal Z}$, we use the expansion of the exponential function in (\ref{Z}) over the electron-phonon interaction term $H_i$, the last addend in the Lagrangian. For brevity, we denote the first two terms in it as ${\cal L}_e$, and the next two addends as ${\cal L}_v$. The result reads
\begin{eqnarray}\label{integrated}
{\cal Z}&=&\int{\cal D}q\exp\bigg(-\int_{0}^{\beta}{\cal L}_v{\rm d}\tau\bigg)\nonumber\\
&&\;\times\int{\cal D}(\xi^*\xi)\exp\bigg(-\int_{0}^{\beta}{\cal L}_e{\rm d}\tau\bigg)
\nonumber\\
&&\;\times\sum_{k=0}^{\infty}\frac{(-1)^k}{k!}\int_{0}^{\beta}{\rm d}\tau_1\ldots{\rm d}\tau_k H_i(\tau_1)\ldots H_i(\tau_k)\nonumber\\
&=&{\rm e}^{-\beta\Omega_e}\int{\cal D}q\exp\bigg\{-\int_{0}^{\beta}{\rm d}\tau\bigg[{\cal L}_v+\sum_{k=0}^\infty\frac{(-1)^k}{(k+1)!}\nonumber\\
&&\;\times\int_{0}^{\beta}{\rm d}\tau_1\ldots{\rm d}\tau_k \widetilde{C}^{(k+1)}(\tau,\tau_1\ldots\tau_k)\bigg]\bigg\}.
\end{eqnarray}
In equation~(\ref{integrated}), $\widetilde{C}^{(k)}(\tau_1\ldots\tau_k)$ are cumulants of the interaction Hamiltonian $H_i$,
\begin{eqnarray*}
&&\widetilde{C}^{(1)}(\tau)=\langle H_i(\tau)\rangle_e, \\
&&\widetilde{C}^{(2)}(\tau_1,\tau_2)=\langle H_i(\tau_1)H_i(\tau_2)\rangle_e-\langle H_i(\tau_1)\rangle_e\langle H_i(\tau_2)\rangle_e, \\
&&\widetilde{C}^{(3)}(\tau_1,\tau_2,\tau_3)=\langle H_i(\tau_1)H_i(\tau_2)H_i(\tau_3) \rangle_e\\
&&\quad-\langle H_i(\tau_1)H_i(\tau_2)\rangle_e\langle H_i(\tau_3)\rangle_e\\
&&\quad-\langle H_i(\tau_1)H_i(\tau_3)\rangle_e\langle H_i(\tau_2)\rangle_e\\
&&\quad-\langle H_i(\tau_2)H_i(\tau_3)\rangle_e\langle H_i(\tau_1)\rangle_e\\
&&\quad+2\langle H_i(\tau_1)\rangle_e\langle H_i(\tau_2)\rangle_e\langle H_i(\tau_3)\rangle_e, \ldots
\end{eqnarray*}
To distinguish these cumulants from the cumulants of electron operators in the previous section, we use tildes over their symbols. The subscript $e$ near angle brackets indicates averaging over electron variables,
$$
\langle{\cal O}\rangle_e={\rm e}^{\beta\Omega_e}\int{\cal D}(\xi^*\xi)\exp\bigg(-\int_{0}^{\beta}{\cal L}_e{\rm d}\tau\bigg){\cal O}.
$$
Hence cumulants in (\ref{integrated}) depend on lattice distortions $q_{\bf k}(\tau)$, which time dependence is related to the continual integration in the last equality in (\ref{integrated}). The average $\langle{\cal O}(\tau_1,\ldots\tau_k)\rangle_e$ can be rewritten in the operator form as $\langle{\cal TO}(\tau_1,\ldots\tau_k)\rangle$, in which operator time dependencies and averaging are determined by the electron part $H$ of the full Hamiltonian. In this form, the cumulants depend parametrically on $q_{\bf k}$. Below, the angle brackets without subscripts denote this type of averaging.

In the exponent in (\ref{integrated}), the first term in the sum describes the shift of lattice vibrations due to the electron-phonon interaction, the second term represents the mode mixing, and subsequent addends the contribution of this interaction to the vibration anharmonism of the third and higher orders. The phonon interactions corresponding to the later terms are carried out by electrons. Therefore, the kernels of these interactions are time-dependent and take into account retardation effects. Terms of the sum in (\ref{integrated}) can be expressed through the cumulants of electron operators considered in the previous section. For our purpose, the first term of the sum is of primary interest. As mentioned above, the shift of the equilibrium positions of lattice vibrations described by this term and related charge inhomogeneity are detected by the quasi-elastic neutron and X-ray scattering \cite{Tranquada88,Birgeneau,Comin}. Besides, as was also indicated, in the crystals with strong electron correlations, the electron-phonon interaction is much smaller than the on-site Coulomb repulsion. Therefore,  in the sum, addends of higher orders are less than the first-order term. In the following discussion, only this term will be taken into account.

In this approximation, the effective Hamiltonian of the phonon subsystem reads
\begin{eqnarray}\label{energy}
H_v&=&\sum_{\bf k}\frac{p_{\bf k}p_{\bf -k}}{2m}+\frac{1}{2}\sum_{\bf ll'}w_{\bf k}q_{\bf k}q_{\bf -k} \nonumber\\
&&+\sum_{\bf kk'\sigma}\sqrt{\frac{2m\omega_{\bf k'}}{N}}\,v_{\bf kk'}\langle a_{\bf k+k',\sigma}^\dagger a_{\bf k\sigma}\rangle\, q_{\bf k'},
\end{eqnarray}
where $p_{\bf k}$ is the vibration momentum. To find shifts of oscillator equilibrium positions we should minimize their potential energy $W$ -- the last two terms in the right-hand side of (\ref{energy}). Recall that the minimum of the free energy in the electron subsystem is achieved in the PS state. Due to this inhomogeneity $\langle a_{\bf k+k',\sigma}^\dagger a_{\bf k\sigma}\rangle\neq\delta_{\bf k',0}\langle a_{\bf k\sigma}^\dagger a_{\bf k\sigma}\rangle$,
\begin{eqnarray}\label{correlator}
\langle a_{\bf k+k',\sigma}^\dagger a_{\bf k\sigma}\rangle&=&\frac{1}{N}\sum_{\bf ll'}{\rm e}^{-{\rm i}{\bf (k+k')l}+{\rm i}{\bf kl'}}\langle a_{\bf l\sigma}^\dagger a_{\bf l'\sigma}\rangle\nonumber \\
&=&\frac{1}{N}\sum_{\bf l}{\rm e}^{-{\rm i}{\bf k'l}}\Big(\langle a_{\bf l\sigma}^\dagger a_{\bf l\sigma}\rangle\nonumber\\
&&\;+\sum_{\bf a}{\rm e}^{{\rm i}{\bf ka}}\langle a_{\bf l\sigma}^\dagger a_{\bf l+a,\sigma}\rangle+\ldots\Big),
\end{eqnarray}
where $\langle a_{\bf l\sigma}^\dagger a_{\bf l\sigma}\rangle$, $\sum_{\bf a}\exp({\rm i}{\bf ka})\langle a_{\bf l\sigma}^\dagger a_{\bf l+a,\sigma}\rangle$, and subsequent electron correlators depend on {\bf l}. Here {\bf a} are vectors connecting nearest neighbors. The correlators $\langle a_{\bf l\sigma}^\dagger a_{\bf l'\sigma}\rangle$ are calculated from Green's function corresponding to the domain, in which sites {\bf l} and ${\bf l'}$ are located,
$$
\langle a_{\bf l\sigma}^\dagger a_{\bf l'\sigma}\rangle=\frac{T}{N}\sum_{{\bf k}j}{\rm e}^{{\rm i}{\bf k(l'-l)}+{\rm i}\omega_j\eta}G({\bf k},j),\quad\eta\rightarrow+0.
$$
They are different in the electron-rich and elec\-t\-ron-poor domain to the same extent as $x_{\bf l}=2\langle a_{\bf l\sigma}^\dagger a_{\bf l\sigma}\rangle$. If {\bf l} and ${\bf l'}$ belong to different domains, the mean between values in the domains can be used for estimating.

Minimizing $W$ in (\ref{energy}) with respect to $q_{\bf k}$, we find
\begin{equation}\label{minimum}
W_{\rm min}=-\frac{1}{N}\sum_{\bf k'}\frac{1}{\omega_{\bf k'}}\bigg|\sum_{\bf k\sigma}v_{\bf kk'}\langle a^\dagger_{\bf k+k',\sigma}a_{\bf k\sigma}\rangle\bigg|^2
\end{equation}
with the value of the minimizing distortions
\begin{equation}\label{qmin}
q_{{\bf k'},{\rm min}}=-\sqrt{\frac{2m\omega_{\bf k'}}{Nw_{\bf k'}^2}}\sum_{\bf k\sigma}v_{\bf k,-k'}\langle a^\dagger_{\bf k-k',\sigma}a_{\bf k\sigma}\rangle.
\end{equation}
Distortions usually associated with the PS correspond to optical phonons with a smooth dependence $\omega_{\bf k'}$ on ${\bf k'}$. Correlators $\langle a_{\bf l\sigma}^\dagger a_{\bf l'\sigma}\rangle$ decrease rapidly with the distance $|{\bf l-l'}|$. Therefore, the dependencies of $\langle a^\dagger_{\bf k+k',\sigma}a_{\bf k\sigma}\rangle$ and $v_{\bf kk'}$ on {\bf k} are also smooth (see (\ref{correlator})). Thus the value of $W_{\rm min}$ is mainly determined by the sum $-\sum_{\bf k'}|v_{\bf \bar{k}k'}x_{\bf k'}|^2$, where ${\bf \bar{k}}$ is some representative value of {\bf k} and $x_{\bf k}$ is the Fourier transform of $x_{\bf l}$. If we suppose that $v_{\bf \bar{k}k'}$ as a function of ${\bf k'}$ has pronounced extrema, the above sum is at a minimum when $x_{\bf k'}$ has maxima at the same values of ${\bf k'}$. Recall that in the used approximations all configurations of electron-rich and electron-poor domains have equal and minimal electron free energy, if they contain $x_rN_r$ and $x_pN_p$ electrons, respectively. Hence the combined free energy of electrons and lattice distortions is at a minimum when the location and widths of electron-rich domains are fitted to the extrema of $v_{\bf \bar{k},l}=N^{-1/2}\sum_{\bf k'}\exp({\rm i}{\bf k'l})v_{\bf kk'}$. As a simple illustration of this statement, let us suppose that $v_{\bf k l}\propto\cos({\bf k}_1{\bf l})$ and $v_{\bf kk'}\propto(\delta_{{\bf k',k}_1}+\delta_{{\bf k',-k}_1})$. Then, for the electron concentration $x_{\bf l}=0.875+0.1\cos({\bf k}_2{\bf l})$, $x_{\bf k'}=0.875\delta_{\bf k',0}+0.05[\delta_{{\bf k',k}_2}+\delta_{{\bf k',-k}_2}]$, $W_{\rm min}\propto-\sum_{\bf k'}|v_{\bf \bar{k}k'}x_{\bf k'}|^2$ is at a minimum when ${\bf k}_1={\bf k}_2$. From (\ref{qmin}), one can see that lattice distortions minimizing $W$ have the same spacial periodicity as quantities $v_{\bf \bar{k},l}$ and $x_{\bf l}$. Thus extrema of the function $v_{\bf kk'}$ of ${\bf k'}$ pick out the most energetically favorable shape and periodicity of electron domains and lattice distortions.

In cuprates, the existence of pronounced maxima in the dependence $v_{\bf kk'}$ on ${\bf k'}$ follows from experimental observations and model calculations. In particular, the out-of-plane oxygen vibrations in cuprates are known to interact strongly with electrons \cite{Barisic,Devereaux}. From the observed structural phase transitions \cite{Axe} and the PS \cite{Tranquada95} in lanthanum cuprates as well as ionic displacements in YBa$_2$Cu$_3$O$_{6.54}$ \cite{Forgan} in the PS state, one can conclude that wavelengths of such vibrations are nearly commensurate with lattice spacings\cite{Hucker}. Longitudinal bond-stretching modes are also presumed to be effective electron scatterers for such wavelengths \cite{Pintschovius,Reznik}. In the mentioned yttrium cuprate, both types of distortions participate in the PS formation \cite{Forgan}.

As follows from the above discussion, symmetry determines the shapes of PS domains and related lattice distortions. In the orthorhombic phase with strong orthorhombicity, $|v_{\bf kk'}|^2$ may have two maxima of equal intensity at ${\bf k'}_{\rm max}$ and $-{\bf k'}_{\rm max}$. They lead to a striped PS with the wavelength $\lambda=2\pi/|{\bf k'}_{\rm max}|$ and the orientation in the direction of ${\bf k'}_{\rm max}/|{\bf k'}_{\rm max}|$. If the orthorhombicity parameter is small or vanishes, $|v_{\bf kk'}|^2$ has four maxima of comparable intensities, and the PS has a checkerboard shape. In the absence of pronounced extrema in the electron-phonon interaction constant, shapes of PS domains become less definite. This case takes place in manganites \cite{Rao}.

From experimental results, it is known that stripes are most stable when their wavelength is commensurate with the lattice spacing \cite{Hucker}. Such stripes are observed at certain levels of doping, e.g., in $p$-type lanthanum cuprates, at the hole concentration $\tilde{x}=1/8$ \cite{Tranquada95}. For other concentrations, stripes are also observed. However, peaks corresponding to them in neutron and X-ray scattering are less intensive \cite{Comin}. Wavelengths of these stripes vary with doping also, which points to the change of extrema positions in $v_{\bf kk'}$ with concentration. Such a dependence was indeed observed in electron-phonon interaction constants calculated by different methods \cite{Devereaux,Banerjee}.

The above electron-phonon Hamiltonian (\ref{ev}) is appropriate, in particular, for the contribution to the PS formation made by Cu-O bond-stretching modes. In the case of out-of-plane oxygen vibrations and one-layer cuprates, this Hamiltonian is inappropriate because it does not take into account the symmetry with respect to the reflection in the Cu-O plane. In accordance with this symmetry, the electron-phonon interaction and phonon potential energy have to contain even powers of oxygen displacements. To ensure the lattice stability, fourth-order terms have to be kept in the potential energy. In lanthanum cuprates, the mentioned displacements are connected with tilts of rigid CuO$_6$ octahedra \cite{Tranquada12}, which can be characterized by the quantity $q_{\bf l}^2=q_{\bf la}^2+q_{\bf lb}^2$. Here $q_{\bf la}$ and $q_{\bf lb}$ describe the octahedron tilt from the {\bf c} crystallographic axis in the {\bf a} and {\bf b} directions. Lanthanum cuprates are $p$-type crystals with doped holes forming Zhang-Rice singlets \cite{Zhang}. Taking into account these facts, the potential energy of the phonon subsystem can be written as
\begin{equation}\label{energy2}
\tilde{W}=\sum_{\bf l}\bigg(-w_2q_{\bf l}^2+\frac{1}{2}w_4q_{\bf l}^4\bigg)-\sum_{\bf ll'}\tilde{v}_{\bf l-l'}\tilde{x}_{\bf l'}q_{\bf l}^2,
\end{equation}
where $\tilde{x}_{\bf l}=\sum_\sigma\langle \tilde{a}^\dagger_{\bf l\sigma}\tilde{a}_{\bf l\sigma}\rangle$ is the concentration of holes in the Zhang-Rice band, $\tilde{a}^\dagger_{\bf l\sigma}$ and $\tilde{a}_{\bf l\sigma}$ are the hole creation and annihilation operators in this band. The hole concentration $\tilde{x}_{\bf l}$ differs in the hole-rich and hole-poor domains in the same manner as the electron concentration above. Minimizing $\tilde{W}$ with respect to $q_{\bf l}$ we find
\begin{eqnarray}\label{minimum2}
\tilde{W}_{\rm min}&=&-\frac{1}{w_4}\sum_{\bf l}\Big(w_2+\sum_{\bf l'}\tilde{v}_{\bf l-l'}\tilde{x}_{\bf l'}\Big)^2\nonumber\\
&=&-\frac{1}{w_4}\Big(Nw_2^2+2\sqrt{N}w_2\tilde{v}_{\bf 0}\tilde{x}_{\bf 0}+\sum_{\bf k}\big|\tilde{v}_{\bf k}\tilde{x}_{\bf k}\big|^2\Big),\nonumber\\
&&
\end{eqnarray}
with minimizing displacements
\begin{equation}\label{qmin2}
q_{{\bf l},{\rm min}}^2=\frac{1}{w_4}\Big(w_2+\sum_{\bf l'}\tilde{v}_{\bf l-l'}\tilde{x}_{\bf l'}\Big).
\end{equation}
In the above equation, $q_{{\bf l},{\rm min}}\neq 0$ if the sum in the brackets is positive and vanishes in the opposite case. In equation (\ref{minimum2}), $\tilde{v}_{\bf k}$ and $\tilde{x}_{\bf k}$ are Fourier transforms of the respective quantities. As in the previously considered case, the potential (\ref{minimum2}) is at a minimum if hole-rich and hole-poor domains are organized in such a manner that pronounced maxima of $|\tilde{x}_{\bf k}|^2$ coincide with those in $|\tilde{v}_{\bf k}|^2$. In other words, the periodicity of the alternating hole-rich and ho\-le-poor domains has to be the same as the periodicity of $\tilde{v}_{\bf l}$ producing the scattering potential generated by lattice distortions. As in the previous case, intensive maxima of $|\tilde{v}_{\bf k}|$ shape domains with different hole concentrations and lattice distortions into stripes or checkerboards.

As shown in \cite{Forgan}, in YBa$_2$Cu$_3$O$_{6.54}$ both oxygen out-of-plane and Cu-O bond-stretching modes contribute to the PS formation. One can assume that this situation is common for cuprate perovskites. Approximating, as above, spatial dependencies of $q_{\bf l}$, $\tilde{v}_{\bf l}$, and $\tilde{x}_{\bf l}$ by trigonometric functions, one can see from (\ref{qmin2}) that the wavelength of $q_{\bf l}$ is twice as many as that of $\tilde{x}_{\bf l}$. It differs from equation~(\ref{qmin}), which shows that in the previous model, these wavelengths are equal. The spacial modulation of stripes in La$_{1.6-x}$ Nd$_{0.4}$Sr$_x$Cu$_2$O$_4$ was determined from the periodicity of the spin ordering. This wavelength was twice as much as that of charge ordering \cite{Tranquada95}. The antiphase alignment of spins near hole-rich domains is most likely connected with the mechanism similar to the Anderson superexchange \cite{Anderson}.

Let us estimate hole concentrations in striped states of lanthanum cuprates at temperatures comparable to those used in the experiments, $(3-60)$~K. We suppose that the one-band Hubbard model is valid for this case and express the temperature in this range in units of $t$. If the exchange constant $J=4t^2/U$ is set to 0.1~eV, the value observed in lanthanum cuprates, and the usually assumed ratio $U=8t$ is valid, we obtain $t=0.2$~eV$\approx 2300$~K. Thus the temperature $T\approx 0.1t$, for which the SCDT results in figure~\ref{Fig2}(a) were calculated, is one-two orders of magnitude higher than those in the mentioned range. With all processes taken into account to obtain this curve, we are unable to attain temperatures of the order of $0.01t$ now. However, we can estimate the lower and upper concentrations in the NEC region using the Hubbard-I approximation. Such calculated dependence $x(\mu)$ is shown in figure~\ref{Fig2}(a) by the blue dotted line. As seen from this curve, in the NEC region, the minimal hole concentration $\tilde{x}_{\rm min}\approx 0.05$ and the maximal one $\tilde{x}_{\rm max}\approx 0.35$. In La$_{2-x-y}$Nd$_y$Sr$_x$Cu$_2$O$_{4+\delta}$, three rows of crystal cells in the stripe form the hole-poor domain, while one row constitutes the hole-rich one \cite{Tranquada95}. Ascribing them the above hole concentrations, for the mean concentration, we obtain $(3\times 0.05+0.35)/4=0.125$, which is consistent with the experiment. In spite of this good agreement, our estimation is rather rough -- apart from using the simplified Hubbard-I approximation, one of the sizes of the hole-rich domain is of the order of the lattice spacing. To apply results obtained in an infinite crystal to domains they have to be much larger.

Our consideration was carried out in the case of the perfect translation symmetry. In this situation, PS structures can freely transfer across the crystal. As known \cite{Tranquada95,Forgan}, such fluctuating stripes can coexist with bulk superconductivity. Real crystals contain different lattice imperfections, which can pin the PS structures. In $p$-type cuprates, such static or quasi-static stripes have wavelengths, which are commensurate with the lattice spacing \cite{Tranquada95,Hucker,Koike}. Other contributory factors for stripe pinning are lattice distortions in the tetragonal low-temperature phase as well as Nd and Zn doping.
The pinning leads to the partial or complete suppression of superconductivity, which points to comparable values of the stripe wavelength and superconducting coherence length.

\section{Concluding remarks}
In this work, we considered the mechanism of phase separation observed in many crystals characterized by pronounced electron correlations. It was related to the regions of negative electron compressibility arising near chemical potentials, which correspond to the change of the ground state in the site Hamiltonians. The existence of these regions leads to the separation of a crystal into electron-rich and electron-poor domains. The separation is accompanied by the energy generation. Therefore, for its implementation, another subsystem is necessary, which absorbs the released energy. In crystals, such a subsystem is phonons. However, their role is not only in the energy absorption but also in the shaping of the electron-rich and electron-poor domains. In the presence of such domains, we considered effective phonon Hamiltonians for two different types of vibrations -- with interaction terms linear and quadratic in lattice distortions. Such Hamiltonians may describe the Cu-O bond stretching and out-of-plane oxygen modes, respectively. In both cases, we found that the minimum of the vibration potential is achieved by the domain arrangement corresponding to pronounced extrema in the electron-phonon interaction constant as the function of phonon momentum. That is, the charge distribution has the wavelength of phonons, which scatter electrons most strongly. In the orthorhombic phase with large orthorhombicity, there are two such modes with opposite momenta. In this case, the charge distribution has the character of stripes, in which the electron density varies along these momenta. In the case of the $C_4$ symmetry, there are four modes with the largest scattering amplitudes, which produces a checkerboard order. In the phase-separated state, equilibrium positions of lattice oscillators vary periodically in space also, being determined by extrema of the electron-phonon interaction constant. Such types of phase separation are inherent, in particular, in cuprates. If the electron-phonon constant has no pronounced extrema, shapes of the electron domains are less definite. This situation is seen in manganites. In all these cases, the phase separation is a consequence of regions of negative electron compressibility caused by strong electron correlations.


\begin{thebibliography}{99}
\bibitem{Tranquada88}J.M. Tranquada, A.H. Moudden, A.I. Goldman, P. Zolliker, D.E. Cox, G. Shirane, S.K. Sinha, D. Vaknin, D.C. Johnston, M.S. Alvarez, A.J. Jacobson, J.T. Lewandowski, J.M. Newsam, Phys. Rev. B {\bf 38}, 2477 (1988)
\bibitem{Birgeneau}R.J. Birgeneau, D.R. Gabbe, H.P. Jenssen, M.A. Kastner, P.J. Picone, T.R. Thurston, G. Shirane, Y. Endoh, M. Sato, K. Yamada, Y. Hidaka, M. Oda, Y. Enomoto, M. Suzuki, T. Murakami, Phys. Rev. B {\bf 38}, 6614 (1988)
\bibitem{Tranquada94}J.M. Tranquada, D.J. Buttrey, V. Sachan, J.E. Lorenzo, Phys. Rev. Lett. {\bf 73}, 1003 (1994).
\bibitem{Tranquada95}J.M. Tranquada, B.J. Sternlieb, J.D. Axe, Y. Nakamura, S. Uchida, Nature {\bf 375}, 561 (1995)
\bibitem{Kremer}R. Kremer, V. Hizhnyakov, E. Sigmund, A. Simon, K.A. M\"uller, Z. Physik B {\bf 94}, 17 (1993).
\bibitem{Hanaguri}T. Hanaguri, C. Lupien, Y. Kohsaka, D.-H. Lee, M. Azuma, M. Takano, H. Takagi, J.C. Davis, Nature {\bf 430}, 1001 (2004)
\bibitem{Vershinin}M. Vershinin, S. Misra, S. Ono, Y. Abe, Y. Ando, A. Yazdani, Science {\bf 303}, 1995 (2004)
\bibitem{Renner}Ch. Renner, G. Aeppli, B.-G. Kim, Yeong-Ah Soh, S.-W. Cheong, Nature {\bf 416}, 518 (2002)
\bibitem{Rao}C.N.R. Rao, P.V. Vanitha, A.K. Cheetham, Chem. Eur. J. {\bf 9}, 828 (2003)
\bibitem{Park}J.T. Park, D.S. Inosov, Ch. Niedermayer, G.L. Sun, D. Haug, N.B. Christensen, R. Dinnebier, A.V. Boris, A. J. Drew, L. Schulz, T. Shapoval, U. Wolff, V. Neu, Xiaoping Yang, C. T. Lin, B. Keimer, V. Hinkov, Phys. Rev. Lett. {\bf 102}, 117006 (2009).
\bibitem{Lang}G. Lang, H.-J. Grafe, D. Paar, F. Hammerath, K. Manthey, G. Behr, J. Werner, B. B\"uchner, Phys. Rev. Lett. {\bf 104}, 097001 (2010)
\bibitem{Shenoy}V.B. Shenoy, C.N.R. Rao, Phil. Trans. R. Soc. A {\bf 366}, 63 (2008)
\bibitem{Pintschovius}L. Pintschovius, W. Reinchardt, M. Kl\"aser, T. Wolf, H. v. L\"ohneysen, Phys. Rev. Lett. {\bf 89}, 037001 (2002)
\bibitem{Reznik}D. Reznik, L. Pintschovius, M. Ito, S. Iikubo, M. Sato, H. Goka, M. Fujita, K. Yamada, G.D. Gu, J.M. Tranquada, Nature {\bf 440}, 1170 (2006)
\bibitem{Tacon}M. Le Tacon, A. Bosak, S.M. Souliou, G. Dellea, T. Loew, R. Heid, K.P. Bohnen, G. Ghiringhelli, M. Krisch, B. Keimer, Nat. Phys. {\bf 10}, 52 (2014)
\bibitem{Hizhnyakov}V. Hizhnyakov, E. Sigmund, Physica C {\bf 156}, 655 (1988)
\bibitem{Machida}K. Machida, Physica C {\bf 158}, 192 (1989)
\bibitem{Zaanen}J. Zaanen, O. Gunnarson, Phys. Rev. B {\bf 40}, 7391 (1989)
\bibitem{Nagaev}E.L. Nagaev, {\em Physics of Magnetic Semiconductors} (Nauka, Moscow, 1979)
\bibitem{Sabczynski}J. Sabczynski, M. Schreiber, A. Sherman, Phys. Rev. B {\bf 48}, 543 (1993)
\bibitem{Moreo}A. Moreo, D. Scalapino, E. Dagotto, Phys. Rev. B {\bf 43}, 11442 (1991)
\bibitem{Seibold}G. Seibold, Eur. Phys. J. B {\bf 35}, 177 (2003)
\bibitem{Sherman08}A. Sherman, M. Schreiber, Phys. Rev. B {\bf 77}, 155117 (2008)
\bibitem{Chia}C. Chia-Chen, Z. Shiwei, Phys. Rev. B {\bf 78}, 165101 (2008)
\bibitem{Heiselberg}H. Heiselberg, Phys. Rev. A {\bf 79}, 063611 (2009)
\bibitem{White}S.R. White, D.J. Scalapino, Phys. Rev. B {\bf 92}, 205112 (2015)
\bibitem{Sherman20a}A. Sherman, Phys. Scr. {\bf 95}, 015806 (2020).
\bibitem{Sherman20b}A. Sherman, Phys. Scr. {\bf 95}, 095804 (2020).
\bibitem{Hubbard}J. Hubbard, Proc. R. Soc. London, Ser. A {\bf 296}, 82 (1966).
\bibitem{Zaitsev}R.O. Zaitsev, Sov. Phys.—JETP, {\bf 43}, 574 (1976)
\bibitem{Izyumov}Yu.A. Izyumov, B.M. Letfulov, J. Phys.: Condens. Matter {\bf 1}, 8905 (1990)
\bibitem{Ovchinnikov}S.G. Ovchinnikov, V.V. Valkov, {\em Hubbard Operators in the Theory of Strongly Correlated Electrons} (Imperial College Press, London, 2004)
\bibitem{Vladimir}M.I. Vladimir, V.A. Moskalenko, Theor. Math. Phys. {\bf 82}, 301 (1990)
\bibitem{Metzner}W. Metzner, Phys. Rev. B {\bf 43}, 8549 (1991)
\bibitem{Pairault}S. Pairault, D. S\'en\'echal, A.-M.S. Tremblay, Eur. Phys. J. B {\bf 16}, 85 (2000)
\bibitem{Sherman18}A. Sherman, J. Phys.: Condens. Matter {\bf 30}, 195601 (2018)
\bibitem{Sherman19a}A. Sherman, Eur. Phys. J. B {\bf 92}, 55 (2019)
\bibitem{Sherman19b}A. Sherman, Phys. Scr. {\bf 94}, 055802 (2019)
\bibitem{Abrikosov}A.A. Abrikosov, L.P. Gor’kov, I.E. Dzyaloshinskii, {\em Methods of Quantum Field Theory in Statistical Physics} (Pergamon, New York, 1965)
\bibitem{Kubo}R. Kubo, J. Phys. Soc. Japan {\bf 17}, 1100 (1962)
\bibitem{Senechal}D. S\'en\'echal, D. Perez, M. Pioro-Ladri\`ere, Phys. Rev. Lett. {\bf 84}, 522 (2000)
\bibitem{Georges}A. Georges, L. de’ Medici, J. Mravlje, Annu. Rev. Condens. Matter Phys. {\bf 4}, 137 (2013)
\bibitem{Balian}R. Balian, C. Bloch, Ann. Phys. (N.Y.) {\bf 60}, 401 (1970)
\bibitem{Sboychakov}A.O. Sboychakov, K.I. Kugel, A.L. Rakhmanov, Phys. Rev. B {\bf 76}, 195113 (2007)
\bibitem{Comin}R. Comin, A. Damascelli, Annu. Rev. Condens. Matter Phys. {\bf 7}, 369 (2016)
\bibitem{Negele}J.W. Negele, H. Orland, {\em Quantum Many-Particle Systems} (Westview Press, New York, 1998)
\bibitem{Barisic}S. Bari\v{s}i\'c, I. Batisti\'c, Europhys. Lett. {\bf 8}, 765 (1989)
\bibitem{Devereaux}T.P. Devereaux, A. Virosztek, A. Zawadowski, Phys. Rev. B {\bf 51}, 505 (1995)
\bibitem{Axe}J.D. Axe, M.K. Crawford, J. Low Temp. Phys. {\bf 95}, 271 (1994)
\bibitem{Forgan}E.M. Forgan, E. Blackburn, A.T. Holmes, A.K.R. Briffa, J. Chang, L. Bouchenoire, S.D. Brown, Ruixing Liang, D. Bonn, W.N. Hardy, N.B. Christensen, M.v. Zimmermann, M. H\"ucker, S.M. Hayden, Nat. Commun. {\bf 6}, 10064 (2015)
\bibitem{Hucker}M. H\"ucker, M.v. Zimmermann, Z.J. Xu, J.S. Wen, G.D. Gu, J.M. Tranquada, Phys. Rev. B {\bf 87}, 014501 (2013)
\bibitem{Banerjee}S. Banerjee, W.A. Atkinson, A.P. Kampf, Phonon-induced emergent charge order in cuprates, arXiv:1907.02618
\bibitem{Tranquada12} J.M. Tranquada, Physica B: Condensed Matter {\bf 407}, 1771 (2012).
\bibitem{Zhang}F.C. Zhang, T.M. Rice, Phys. Rev. B {\bf 37}, 3759 (1988)
\bibitem{Anderson}P.W. Anderson, Phys. Rev. {\bf 79}, 350 (1950)
\bibitem{Koike}Y. Koike, M. Akoshima, M. Aoyama, K. Nishimaki, T. Kawamata, T. Adachi, T. Noji, M. Kato, I. Watanabe, S. Ohira, W. Higemoto, K. Nagamine, H. Kimura, K. Hirota, K. Yamada, Y. Endoh, Physica C {\bf 357-360}, 82 (2001)
\end{thebibliography}
\end{document}